\documentclass[10pt]{article}
\usepackage{amsmath,amsfonts,amssymb,amsthm,bbm,bm}
\usepackage{pdfsync}
\usepackage{float}
\usepackage{subfigure}
\usepackage{graphicx}
\usepackage[latin1]{inputenc}
\usepackage{hyperref}
\usepackage[all]{xy}
\usepackage{stmaryrd}
\usepackage{psfrag}
\usepackage{rotating}
\usepackage{epstopdf}
\usepackage{braket}
\usepackage{amsthm}

\newcommand{\vgraph}{\mathfrak{n}}

\newcommand{\bit}{\begin{itemize}}
\newcommand{\eit}{\end{itemize}}
\newcommand{\bd}{\begin{description}}
\newcommand{\ed}{\end{description}}

\newcommand{\bc}{\begin{center}}
\newcommand{\ec}{\end{center}}

\newcommand{\be}{\begin{equation}}
\newcommand{\ee}{\end{equation}}
\newcommand{\bea}{\begin{eqnarray}}
\newcommand{\eea}{\end{eqnarray}}
\newcommand{\bs}{\begin{subequations}}
\newcommand{\es}{\end{subequations}}

\begin{document}

\title{\bf Quantum reduced loop gravity effective Hamiltonians from a statistical regularization scheme}

\author{\Large{Emanuele Alesci$^a$, Gioele Botta$^b$ and Gabriele V. Stagno$^{c,d}$}
\smallskip \\ \small{$^{a,b}$SISSA, Via Bonomea 265, I-34136 Trieste, Italy, EU and INFN, Sez. di Trieste.} \\  
\small{$^c$Sapienza University of Rome, P.le Aldo Moro 5, (00185) Roma, Italy }\\
\small{$^d$Aix Marseille Univ., Univ. de Toulon, CNRS, CPT, UMR 7332, 13288 Marseille, France}\\ \\ \small{Email: $^a$emanuele.alesci@sissa.it, $^b$gioele.botta01@universitadipavia.it, $^{c,d}$gabriele.stagno@uniroma1.it}
}
\date{\today}

\maketitle

\begin{abstract}
\noindent
We introduce a new regularization scheme for Quantum Cosmology in Loop Quantum Gravity (LQG) using the tools of Quantum Reduced Loop Gravity (QRLG). It is obtained considering density matrices for superposition of graphs based on statistical countings of microstates compatible with macroscopic configurations. We call this procedure statistical regularization scheme. In particular, we show how the $\mu_0$ and $\bar{\mu}$ schemes introduced in Loop Quantum Cosmology (LQC) emerge with specific choices of density matrices. Within this new scheme we compute effective Hamiltonians suitable to describe quantum corrected Friedmann and Bianchi $I$ universes and their leading orders coincide with the corresponding effective LQC Hamiltonians in the $\bar{\mu}$ scheme. We compute the next to the leading orders corrections and numerical investigation of the resulting dynamics shows evidence for the emergent-bouncing universe scenario to be a general property of the isotropic sector of QRLG.
\end{abstract}

\tableofcontents
\section{Introduction}

Loop Quantum Gravity (LQG) \cite{Rovelli:2004tv,Thiemann:2007zz,Ashtekar:2004eh,Dona:2010hm} is one of the main proposals towards a non perturbative quantization of the gravitational field. The theory describes the gravitational field through quanta of spacetime dual to spin networks, quantum states of geometry labelled by graphs colored by SU(2) quantum numbers, which form a basis for the LQG kinematical Hilbert space. If the kinematical space is well defined, finding the dynamical one still remains an open issue. This fact should not be surprising at all since already at the classical level only few solutions of the Einstein equations are known. It is only in presence of symmetries that the Einstein equations become manageable and explicit solutions can be found, as for cosmology where one can consider diagonal metrics and study homogeneous spacetimes neglecting their spatial dependence. This procedure can be used to simplify also the quantization procedure. 

Loop Quantum Cosmology (LQC) is the first model able to give a consistent quantum version of homogeneous spacetimes in the LQG framework \cite{Ashtekar:2011ni,Bojowald:2011zzb,Agullo:2016tjh}. LQC is built imposing the cosmological symmetries at the classical level and the resulting finite dimensional system is then quantized using LQG machinery. However, the classical symmetry reduction followed by quantization loses relevant structures of the original theory. Indeed, consistency for LQC requires the so called $\bar{\mu}$ scheme \cite{Ashtekar:2006uz,Ashtekar:2006wn} where several input from the full theory need to be imported, and within which a resolution of the classical Big-Bang singularity with a Big-Bounce and a power spectrum for the cosmic microwave background in agreement with what we observe today are found \cite{Agullo:2012fc,Agullo:2013ai,Ashtekar:2015dja}.

In the last few years Quantum Reduced Loop Gravity (QRLG) has been proposed as an approach to avoid the loss of foundamental LQG structures when dealing with symmetry reduced systems \cite{Alesci:2012md,Alesci:2013xd,Alesci:2014uha,Alesci:2015nja,Alesci:2016gub,Cianfrani:2015oha}. In QRLG the symmetry reduction is performed in two steps, first imposing a gauge fixing and then considering the symmetry reduction. The gauge fixing conditions are imposed on the kinematical Hilbert space of the full theory directly on the states, weakly \cite{Alesci:2013xya}. In the case of diagonal gauge the resulting gauged fixed Hilbert space is made of cuboidal graphs colored by $U(1)$ quantum numbers, which brings a drastic simplification in computing matrix elements of the kinematical operators (for example the volume operator is diagonal). 

In LQC the so called regularization schemes are fundamental to define a consistent semiclassical dynamics, and can be fixed importing two key features of the full theory: the discretization of the geometrical operators and the graph structure that are lost after the symmetry reduction at the classical level. In QRLG the situation is completely different since both emerge naturally during the quantization, and both the LQC regularization schemes can be recovered in the Friedmann case \cite{Alesci:2016gub,Alesci:2016rmn}. 

In this paper we show how to recover these two schemes in general through a suitable choice of density  matrices. However, in both cases one has to consider the total number of nodes $N$ that characterizes a graph as a classical number, i.e. has to neglect the fluctuations associated to $N$. Indeed, we show how the regularization scheme presented for the first time in \cite{Alesci:2016rmn} is more general than the LQC schemes since it arises from statistical arguments which take into account the fluctuations associated to $N$. We name this scheme statistical regularization. The fluctuations associated to $N$ are the one responsible for the drastic departure from the LQC bouncing scenario, leading to the so-called emergent bouncing universe \cite{Alesci:2016xqa}. 

The improved regularization can be recovered also for Bianchi $I$ spacetime through a suitable statistical counting, furthermore, a new effective Hamiltonian for a Bianchi $I$ spacetime is obtained: it coincides at the leading order with the one of LQC but presents new corrections. The counting needed for Bianchi $I$ provides a unified framework for cosmology: it is possible to consider the Friedmann density matrix as the reduced density matrix that characterizes Bianchi $I$.

First we start by reviewing basis of LQC and QRLG, then in section \ref{stat} we introduce the statistical regularization scheme showing, as a concrete example, how the volume counting leads to the effective dynamics for FLRW previously found in \cite{Alesci:2016xqa}. In section \ref{Bianchi} we use a new counting, the area counting, which allows to deal with the non isotropic case and we compute the effective Hamiltonian for Bianchi $I$, in section \ref{AREAFRID} we apply the area counting to the FLRW case, showing how the resulting state derives from the Bianchi $I$ reduced density matrix in section \ref{REDUCED}, and studying its associated dynamics in section \ref{result}. The last section is devoted to conclusions. Throughout the paper we adopt units $G$=$\hbar$=$c$=$1\,$.

\section{Loop Quantum Cosmology}

The quantization procedure carried out by LQC is a minisuperspace quantization whose starting point is the phase space of the ADM formulation of general relativity \cite{Deser:1960zzc} where the desired cosmological symmetries are imposed at the classical level. Mimicking the LQG procedure, the reduced phase space in terms of fluxes and holonomies is then quantized with the tools of polymer quantization \cite{Halvorson,Ashtekar:2002sn,Corichi:2007tf}.

Below we will briefly review the LQC quantization of the (spatially flat) FLRW geometry filled by a massless scalar field and a vanishing cosmological constant.

\subsection{LQC quantization of $k=0$ FLRW spacetime sourced by a scalar field}

The classical FLRW, $k=0$ geometry is captured by the following line element
\be
ds^{2}=-dt^{2}+a^{2}(t)(dx^{2}+dy^{2}+dz^{2})\,, \label{metf}
\ee
where $a(t)$ is the scale factor and $t$ the cosmological time (we choose the lapse function equal to $1$). The Hamiltonian formulation of this spacetime requires the introduction of a fiducial cell $V_{0}$, otherwise, the homogeneity of the model leads to a divergence in the spatial integration needed to define the Lagrangian, the Hamiltonian and the symplectic structure. $V_{0}\,,$ in a non compact topology like $\mathbb{R}^{3}$, should be thought as an infrared regulator whose choice must not influence the physical results obtained sending $V_{0}\to \infty$. 

The phase space for a FLRW geometry filled by a massless scalar field $\phi$ can be parametrized by the quadruplet $(a,P_{a},\phi,P_{\phi})$, where $a$ and $P_{a}=-a\dot{a}$ are the conjugate variables for the geometry and $\phi$ and $P_{\phi}=V\dot{\phi}$ for the scalar field, where $V=a^{3}V_{0}$ represents the physical volume of a region of the universe. 

Due to the high symmetry of the model, it turns out that the spatial diffeomorphism constraint is automatically satisfied and the only constraint we are left with is the Hamiltonian constraint: 
\be
H^{FLRW}_{LQC}=-\frac{3}{8\pi}\frac{V_{0}P^{2}_{a}}{a}+\frac{P^{2}_{\phi}}{2V_{0}a^{3}}=0
\ee
and the only non vanishing Poisson brackets are
\be
\{a,P_{a}\}=\frac{4\pi}{3V_0}\,,\quad \{\phi,P_{\phi}\}=1\,.
\ee
The starting point of LQG quantization is to cast General Relativity in a form similar to the one of a gauge theory, this is done introducing a new pair of canonical variables: the Ashtekar variables \cite{Ashtekar:1986yd}. These new pair are the (Ashtekar) connection $A_{a}^{i}$ and its conjugate momentum $E^{a}_{i}$, the densitized triad field, which is a Lie-algebra valued vector field of density weight one. This is also the starting point of the LQC quantization, where these pair of canonical variables can be written (thanks to the symmetries of the FLRW $k=0$ model) in the following reduced form:
\be
A_{a}^{i}=c(t)V_{0}^{-1/3}\delta_{a}^{i}\,, \ \ \ E_{i}^{a}=p(t)V_{0}^{-2/3}\delta_{i}^{a}\,, \label{lqcvar1}
\ee
where $\delta_{a}^{i}$, $\delta_{i}^{a}$ are respectively a set of orthonormal co-triad and triad adapted to the edges of the fiducial cell $V_{0}$. From \eqref{lqcvar1} it is clear that in such a reduced model all the non trivial information is contained in the couple $(c(t),p(t))$, which is related to the scale factor and its time derivative as follows:
\be
c=V_{0}^{1/3}\gamma\dot{a}\,, \ \ \ p=a^{2}V_{0}^{2/3}\,,
\ee
where $\gamma\approx 0.24$ is the Immirzi parameter whose value is fixed by the black-hole entropy calculation \cite{Ashtekar:1997yu} and their Poisson brackets are
\be
\{c,p\}=\frac{8\pi\gamma}{3}\,.
\ee

To pass to the quantum theory we have first to specify the variables which have to be promoted to operators. Always following the LQG quantization procedure in LQC there is a natural choice: holonomies and fluxes. Again a simplification due to homogeneity and isotropy arises since it suffices to consider holonomies along edges of the fiducial cell and fluxes across faces $S$ of $V_{0}$ and these two functions take respectively the following form:
\be
h_{\mu}(c)=e^{i\mu c/2}\,, \ \ \ \ E(S)=p\,,
\ee
where $\mu$ is the ratio between the coordinate lenght of a path parallel to an edge of the fiducial cell $V_{0}$ and the lenght of the edge itself. Working in connection representation the LQC Hilbert space is spanned by almost periodic functions,
\be
\Psi(c)=\sum_{n}\alpha_{n}\,e^{i\mu_{n}c/2}\,,
\ee
which live in the space of square integrable functions over the Bohr compactification of the real line $L^{2}(\overline{\mathbb{{R}}},dc)\,,$ whose square module is defined as
\be
|\Psi|^{2}:=\lim_{D\to\infty}\frac{1}{2D}\int_{-D}^{+D} dc\, \overline{\Psi}(c)\Psi(c)=\sum_{n}|\alpha_{n}|^{2}\,.
\ee 
An orthonormal basis of this Hilbert space is given by $\langle c|h_{\mu}\rangle=h_{\mu}(c)=e^{i\mu c/2}\,,$ since they satisfy
\be
\langle h_{\mu}|h_{\mu'}\rangle=\delta_{\mu \mu'}\,.
\ee

The next step is to promote holonomies and fluxes to operators. In the ``holonomy representation'', the holonomy operator acts by multiplication and its momentum, i.e. the flux operator, as a derivative:
\be
\hat{h}_{\mu}\Psi(c)=e^{i\mu c/2}\Psi(c)\,, \ \ \ \ \hat{p}\Psi(c)=-i\frac{8\pi\gamma l_P^{2}}{3}\frac{d\Psi}{dc}\,,
\ee
where $l_P$ is the Planck lenght. 

Given the self-adjointness of $\hat{p}$ it is often convenient to work on a representation where it acts diagonally:
\be
\hat{h}_{\mu}\Psi(\mu')=\Psi(\mu+\mu')\,, \ \ \ \hat{p}\Psi(\mu)=\frac{4\pi\gamma l_P^{2}}{3}\mu\,\Psi(\mu)\,,\label{LQCoperators}
\ee
 
and this is enough to define the kinematical space of LQC. 
 
Dynamics needs a suitable operator-promotion of the Hamiltonian constraint, which in this simplified model reads:
 \be
 C_{grav}(c,p)=-\gamma^{-2}V_{0}^{-1/3}\epsilon^{ij}_{k}\delta_{i}^{a}\delta_{j}^{b}|p|^{2}F_{ab}^{k}(c)\,, \label{Hquanlqc}
 \ee
 where $F_{ab}^{k}$ is the field strenght. 
The LQC reduced constraint has a direct dependence on the connection $c$ which has no operator analogous in the Hilbert space of LQC. As we will see in the next subsections, a regularization of $C_{grav}$ using the strategy adopted in full LQG defines $F_{ab}^{k}$ in terms of holonomies and a suitable definition of $\mu$. However, the choice of $\mu$, i.e. the choice of a particular regularization scheme, deeply influences the dynamics of the model, as we are ready to discuss in the next section.

\subsection{Regularization schemes in LQC}

To understand the physical meaning of the QRLG Hamiltonian obtained in \cite{Alesci:2016rmn} it is useful to review the semiclassical dynamics of LQC. As we have seen in the previous section, the quantization procedure in LQC has been carried out imposing homogeneity and isotropy already at the classical level. This implies that relevant features of the full theory as the graph structure of the kinematical states and the explicit discretization of the geometric operators are lost. However they can be imported from the full theory through the so called regularization schemes. 
 
Let us consider the LQC effective Hamiltonian for a flat FLRW model in the presence of a massless scalar field \cite{Ashtekar:2011ni}:
\be
H^{FLRW}_{LQC}=-\frac{3}{8\pi\gamma}\sqrt{p}\,\frac{\sin^{2}(\mu c)}{\mu^{2}}+\frac{P_{\phi}^{2}}{2 p^{3/2}}=0\,, \label{hamlqc}
\ee
the choice of a regularization scheme is directly related to the definition of $\mu$ that appears in \eqref{hamlqc} and it has a deep impact on the dynamics of the universe; so far in the literature two different definitions for $\mu$ have been proposed, the so called $\mu_{0}$ \cite{Ashtekar:2006uz} and $\bar{\mu}$ scheme \cite{Ashtekar:2006wn}. 

\subsection{$\mu_{0}$ scheme}

As we have seen, the Hamiltonian constraint \eqref{Hquanlqc} depends on $F_{ab}^{k}\,$, which in turn has terms proportional to $c^{2}$ \cite{Ashtekar:2003hd} and then it cannot be promoted to a well defined operator in LQC Hilbert space. There is an easy way to overcame this difficulty:  writing the $ab$ component of $F_{ab}^{k}$ in terms of holonomies around a plaquette in the $a-b$ plane. Thanks to the spatial homogeneity and isotropy it suffices to consider one square plaquette of the fiducial cell $V_{0}\,,\Box_{ij}\,,$ where $i$ and $j$ denote two of its boundary edges, and $F_{ab}^{k}$ takes the following form:
\be
F_{ab}^{k}=2 \lim_{Ar_{\Box\to 0}} Tr\bigg(\frac{h_{\Box_{ij}}-\mathbb{I}}{Ar_{\Box}}\tau^{k}\bigg)\delta^{i}_{a}\delta^{j}_{b}\,, \label{curvhol}
\ee
where $Ar_{\Box}=(\mu_{0} V_{0})^{2/3}$ is the area of the plaquette and $\tau^{k}=-i\sigma^{k}/2\,,$ where $\sigma^{k}$ is the Pauli matrix in the $k$ direction. However, the limit defined in \eqref{curvhol} cannot be really taken because the weak continuity with respect to $\mu_{0}$ fails. This issue is solved taking into account one of the fundamental features of the full theory: the quantized nature of the geometry. 

LQC takes advantage of the fact that in LQG there is a well defined area operator $\hat{Ar}_{LQG}$ for a graph $\Gamma$, with a discrete spectrum
\be
\hat{Ar}_{LQG}\,|\Gamma,j,i>= 8\pi\gamma l_P^{2}\sqrt{j(j+1)}\,|\Gamma,j,i>, \label{arlqg}
\ee
where $|\Gamma,j,i>$ denotes an element of the spin-network basis:$\,j$ is a quantum number associated to each link of $\Gamma$ and labels the irreducible representations of SU($2$) (whose nonvanishing minimum is fixed by $j=1/2$), $i$ labels the intertwiners at each node of $\Gamma\,$. In this sense the non existence of the limit \eqref{curvhol} is directly related to the discretization of the geometrical operators, hence a reasonable request should be to shrink the plaquette till its surface equals the minimum non vanishing eigenvalue of the LQG area operator. Let us consider then an holonomy along an edge of this minimal plaquette, which we denote as $h_{\mu_{0}}=e^{i\mu_{0} c/2}\,$, from (\ref{LQCoperators}) we have
\be
\hat{p}_{LQC}\,h_{\mu_{0}}=\frac{4\pi\gamma l_P^{2}}{3}\mu_{0}h_{\mu_{0}} \label{mu0lqc}
\ee
and the value of $\mu_{0}$ is determined equating the spectra of the area operator in \eqref{mu0lqc} with the one in \eqref{arlqg}, for $j=1/2$ we find 
\begin{equation}
\mu_{0}=3\sqrt{3}\,.\label{3rad3}
\end{equation}

However, this choice has unwanted physical consequences. Considering the semiclassical LQC Hamiltonian \eqref{hamlqc} and taking for $\mu$ the constant value (\ref{3rad3}), the following modified Friedmann equation \cite{Ashtekar:2006uz} follows
\be
\bigg(\frac{\dot{a}}{a}\bigg)^2=\frac{8\pi}{3}\rho\bigg(1-\frac{\rho}{\rho_{cr}}\bigg)\,,\label{LQCfriedeq}
\ee
which tells us that when the density $\rho=P_{\phi}^{2}/2p^{3}$ of the universe reaches the critical value $\rho_{cr}$ the big bang singularity is replaced by a bounce. In this regularization scheme $\rho_{cr}= 2^{1/3}3/((8\pi\gamma^{2}\Delta_{LQG} )^{3/2}P_{\phi})$ \cite{Ashtekar:2006uz}, where $\Delta_{LQG}$ is the minimum area eigenvalue in LQG, and this poses an issue in the consistency of the model: in classical general relativity the Friedmann equations are insensitive to the spatial topology, indeed in a FLRW model with $k=0$ and $\Lambda=0$ having a non compact topology like $\mathbb{R}^{3}$ or a compact one like $\mathbb{T}^{3}$ gives always the same equations for the scale factor. But as we have seen, spatially homogeneity requires the introduction of an infrared regulator $V_{0}$ from which $\rho_{cr}$ happens to depend since $P_{\phi}=p^{3/2}\dot{\phi}$, where $p^{3/2}=a^{3}(t)V_{0}$, and for non compact spatial topology taking the limit $V_{0}\to\infty$ leads to the undesired feature of a vanishing $\rho_{cr}\,$.

A possibility to overcome this inconsistency would be considering the topology of $\mathbb{T}^{3}$, indeed in this case the regulator is not needed having a closed spacetime. Nonetheless, also in this case the $\mu_{0}$ scheme fails to provide a consistent description of the universe because the value of $\rho_{cr}$ at which the bounce occurs is too far away from the Planck scale by several orders of magnitude, being $\,\rho_{cr}=10^{-32}g/cm^{3}$ \cite{Ashtekar:2011ni}.
 
These inconsistencies are solved within the modern regularization scheme in LQC, the $\bar{\mu}$ scheme.

\subsection{$\bar{\mu}$ scheme}

The $\bar{\mu}$ scheme borrows another ingredient from the full theory: the graph structure of the kinematical states. A given $\hat{p}_{LQC}$ eigenstate $\Psi(c)=e^{i\mu c/2}\,,$ represents the physical geometry of a face of $V_{0}$ with area $(4\pi\gamma l_P^{2}/3)\mu$. Heuristically, it is possible to look for a correspondence between LQC states and LQG spin networks: a suitable spin network that can describe the geometry of an homogeneous and isotropic cosmology should have edges parallel to the three directions to which the cubical cell $V_{0}$ is adapted. Considering a face of the fiducial cell, the global area is dual to a spin-network $|\Gamma,j,i\rangle$ with $N$ edges piercing orthogonally the considered face, each of them carrying the same quantum number $j$ because of the requested homogeneity and isotropy. The area operator on this state has the following action:
\be
\hat{Ar}_{LQG}\,|\Gamma,j,i>_{FLRW}=8\pi\gamma l_P^{2}N\sqrt{j(j+1)}\,|\Gamma,j,i>_{FLRW}\equiv p\,|\Gamma,j,i>_{FLRW}\,.\label{alqgfrw}
\ee
In order to achieve the best coarse grained homogeneity, one should maximize the number of edges and minimizing the dual areas, i.e. choose a spin network whose every single edge carries the minimal spin $j=1/2$. Every face of $V_{0}$ is then seen as built up by a collection of elementary plaquettes whose physical area is 
\be
\Delta_{LQG}=4\sqrt{3}\pi\gamma l_P^{2}\label{deltaLQG}
\ee
and the total fiducial area $V_0^{2/3}$ of a single face must satisfy 
\be
 V_{0}^{2/3}=N(\bar{\mu}V_{0}^{1/3})^{2}, \label{mua}
\ee 
where $\bar{\mu}=l_{0}/(V_{0}^{1/3})$ and $l_{0}$ is the coordinate lenght of an elementary edge.  From \eqref{mua} and \eqref{alqgfrw}  we find respectively 
\be
\bar{\mu}^{2}=1/N
\ee
and
\be
\bar{\mu}^{2}p=\Delta_{LQG}\,.\label{mu2}
\ee
The latter defines the $\bar{\mu}$ scheme. Using (\ref{mu2}) in \eqref{hamlqc} one gets the modified Friedmann equation \eqref{LQCfriedeq} with the difference that within this scheme the critical density turns out to be $\rho_{cr}=3/(8\pi\gamma^{2}\Delta_{LQG})=0.41\rho_P\,$, that is, a (Planckian) value that does not depend on the regulator $V_0$.

\subsection{Summary and comments about the two schemes in the light of QRLG}

As we have seen, LQC requires regularizations from which the resulting dynamics is deeply affected. Furthermore, both the two regularizations available require the help of the full theory, i.e. LQG. 

In the $\mu_{0}$ scheme the fundamental ingredient imported from LQG is the discretized nature of geometry, which relates the value of $\mu_{0}$ to the minimum non vanishing eigenvalue of the area operator. The corresponding physical picture is that of a spin network associated to a graph with just one single edge piercing orthogonally a face of $V_{0}$ and then, during the expansion of the universe, the only quantum number that grows is $j$. The consequences of this choice leads to an unwanted result, indeed the critical density at which the bounce occurs is not fixed but inversely proportional to $P_{\phi}$.  

The $\bar{\mu}$ scheme can be seen as an extension of the previous scheme since not only the quantized geometry, but even the graph structure of full LQG is taken into account. Here the spin network representing the state of the quantum FLRW geometry is associated to a graph with $N$ edges and the regulator is inversely proportional to this number: $\bar{\mu}^{2}p=\Delta_{LQG}$. In this picture, during the expansion of the universe there is a growth of the total number $N$ of plaquettes pierced by edges which all carry the fixed minimum spin $j=1/2$. 

What is missing in both schemes is a possible contribution by a superposition of different spin network states, i.e. states labelled by different value of $N$ \textit{and} $j$, collectively describing the same geometry. At the quantum level it is indeed quite reasonable to expect the effective Hamiltonian to take contributions also from these ``less coarse-grained'' states. We will see that this emerges naturally in the context of QRLG, with a new regularization scheme that can be seen as an extension of the LQC-ones, including as particular cases both the $\mu_{0}$ and the $\bar{\mu}$ scheme. Furthermore, since QRLG inherits in a natural way from the full theory the graph structure and the discretization of geometrical operators, regularizations here do not require any external input from the full theory.

\section{Quantum Reduced Loop Gravity}

QRLG is a gauge fixed version of LQG which implements at the quantum level a reduction of degrees of freedom according to the symmetry of the system. This approach turns out to be particulary fruitful when addressed to cosmology, being enough here to select as an evolution operator only the part of the full Hamiltonian constraint which generates the evolution of the homogeneous part of the metric, i.e. the euclidian constraint, and to address quantum non homogeneous systems via a perturbative expansion from the homogeneous configuration.

\subsection{Kinematics}

Starting from the kinematical Hilbert space of LQG, QRLG implements the following gauge fixing conditions to restrict to a diagonal 3-metric tensor $\eta^{ab}(t,x)$ and triad fields $E^{a}_i(t,x)\,$ : 

\be
\eta^{ab}(t,x)=\delta_{ij}E^a_i(t,x)E^b_j(t,x)=0\mbox{\quad for\,}a\neq b\,,\quad \qquad \chi_i=\epsilon_{ij}^{\phantom{12}k}\,E^a_k(t,x)\,\delta^j_a = 0\,.\label{dtgf}
\ee
Mimicking the spinfoam procedure \cite{Engle:2007qf}, these conditions are promoted to operators \cite{Alesci:2013xya} and implemented \textit{weakly}, {\it i.e.}  used to select the subspace of quantum states $|\phi_{\alpha}\rangle$ such that
\be
\langle \phi_{\alpha} | \hat{\eta}^{ab} | \phi_{\beta} \rangle = 0\,, \qquad \langle \phi_{\alpha} | \hat{\chi}_i | \phi_{\beta} \rangle = 0\,,\label{weakgauge}
\ee 
and it turns out that these conditions hold if:
\begin{enumerate}
{\item the graphs have links only along fiducial directions, i.e. are \textit{cuboidal}\,;}
{\item group elements belong to the $U(1)$ subgroup obtained by stabilizing $SU(2)$ to the chosen fiducial direction $\delta_i=\delta_i^a\partial_a$ along which the link belongs\,.}
\end{enumerate}
The first condition is fullfilled by $SU(2)$ spin networks associated to cuboidal graphs $\Gamma$ and taking as admissible diffeomorphims only transformations that preserve the cuboidal structure, i.e. only redefinitions of fiducial coordinates:
\be
x'_1=x_1'(x_1)\,,\quad x_2'=x_2'(x_2)\,,\quad x_3'=x_3'(x_3)\,.\label{reddiff}
\ee 
The second condition is realized at each link $l$ by projecting the magnetic indexes of the Wigner matrices on the $SU(2)$ coherent states  that have maximum or minimum magnetic number along the fiducial direction $\vec{u}_l$, i.e. projecting on $|\pm j,\vec{u}_l \rangle:=D^{j_l}(\vec{u}_l)|j,m=\pm j\rangle$, where $D^{j_l}(\vec{u}_l)$ is the matrix the rotates the $z$ direction onto $\vec{u}_l\,$. Hence, a generic basis element of the kinematical Hilbert space of QRLG is  written as 
\be
 \langle h | \Gamma, \{ m_l\}, \{x_n\} \rangle = \prod_{n\in\Gamma}\langle\{j_{l}\}, \{ x_n\}|\{m_{l}\},  \{ {\vec{u}}_l\} \rangle
\prod_{l} \;{}^l\!D^{j_{l}}_{m_{l} m_{l}}(h_{l}),\,\quad m_l=\pm j_l\,, \label{base finale}
\ee
where ${}^l\!D^{j_{l}}_{m_{l} m_{l}}(h_{l}):=\langle m_l, \vec{u}_l | D^{j_l}(h_l)|m_l,\vec{u}_l\rangle$  and the coefficients $\langle\{j_{l}\}, \{ x_n\}|\{m_{l}\},  \{ {\vec{u}}_l\} \rangle$ are one-dimensional intertwiners given by the projection of Livine-Speziale coherent intertwiners \cite{Livine:2007vk} on the standard $SU(2)$ intertwiner basis. QRLG-holonomies are associated to a given direction $k$ along which the link $l$ seats, i.e.
\be 
 h_l=P e^{i\int_l A^{k}\tau_{k}}:=e^{i\theta_l m_l}\,\qquad \mbox{$k,l\,$ not summed\,,}\label{QRLGholo}
 \ee
thus the action of the corresponding holonomy operators follows simply from $U(1)$ recoupling theory.

Flux operators are given by projecting those of LQG down to QRLG by means of (\ref{weakgauge}), with the result that the only non vanishing QRLG-fluxes are those corrisponding to surfaces orthogonal to the triad, i.e. $E_i(S^j)\neq 0 \Leftrightarrow i=j$, and their action reads
\be
\hat{E}_i(S^i)\,{}^l\!D^{j_{l}}_{m_{l} m_{l}}(h_{l})= 8\pi\gamma l_P^2\, m_{l}\,{}^l\!D^{j_{l}}_{m_{l} m_{l}}(h_{l}) \qquad l_i\cap S^i\neq \oslash\,.\label{redei}
\ee
Fluxes coincide with the areas pierced by the links and we can define them as area operators $\hat{p}_i$ in the $i$ direction, then from (\ref{redei}) we read that the QRLG area gap $\Delta$, i.e. the minimal (positive) eigenvalue for $\hat{E}_i$, is  
\begin{equation}
\Delta=4\pi\gamma l_P^2=\frac{\Delta_{LQG}}{\sqrt{3}}\,.\label{deltaQRLG}
\end{equation}
Out of these fluxes we easily construct the volume operator $\hat{V}=\int d^3x \sqrt{|\hat{E}_1(S^1)\hat{E}_2(S^2)\hat{E}_3(S^3)|}$ which is diagonal and then allows to perform easier analytical computations.

\subsection{Dynamics}

The kinematical arena of QRLG is just gauge-fixed LQG. The phase space reduction is implemented dynamically, i.e. considering only that part of the scalar constraint which generates the evolution of the homogeneous part of the metric and choosing coherent states to describe homogeneous states.

In the homogeneous case, the spatial part of the metric has each scale factor depending on time and its associated coordinate (because of the residual symmetry (\ref{reddiff})) i.e.
\be
ds^2=dt^2-a^2_1(t,x_1)dx_1^2-a^2_2(t,x_2)dx_2^2-a^2_3(t,x_3)dx_3^2\,,
\ee
for which spin connections vanish and Ashtekar connections are diagonal. The original $SU(2)$ Gauss constraint reduces to three independent $U(1)$ ones and the vector constraint vanishes identically. Finally, the last constraint benefits of a big simplification, since substituting the diagonal form for connections and momenta into the scalar constraint of full LQG the Lorentzian part turns out to be proportional to the Euclidean part. On the contrary, non homogeneous configurations are characterized by scale factors $a_i$ which now are generic functions of \textit{all} fiducial coordinates and because of that neither Ashtekar connections are diagonal nor the vector constraint vanishes identically and the Hamiltonian analysis is very complicated (see for instance \cite{Bodendorfer:2014vea}). 

Effective dynamics can be achieved in QRLG taking advantage of the coherent state technology developed for full LQG \cite{Thiemann:2000bw,Thiemann:2002vj} which allows a definition of the following coherent states \cite{Alesci:2014uha}:
\begin{equation}
\psi^{{\bf\alpha}}_{\Gamma\{H'\}}=\sum_{\{m_{l}\}}\prod_{n\in\Gamma} \langle\{j_{l}\}, \{x_n\}|\{m_{l}\},  \{\vec{u}_l\} \rangle^*\;\prod_{l\in\Gamma} \psi^\alpha_{H'_{l}}(m_{l})\;\langle h|\Gamma, \{m_l\}, \{x_\vgraph \}\rangle\,,
\label{semiclassici ridotti inv}
\end{equation}
where the coefficients $\psi^\alpha_{H'_l}(m_{l})$ are given by
\begin{equation}
\psi^\alpha_{H'_l}(m_{l})=(2j_{l}+1)e^{-j_{l}(j_{l}+1)\frac{\alpha}{2}}e^{i\bar{\theta}_l m_{l}}e^{\frac{\alpha}{8\pi\gamma l_P^2}E'_i m_{l}}\,,\quad j_l=|m_l|\,.
\end{equation}
States (\ref{semiclassici ridotti inv}) are labelled by the variables $H'_l:=h^{\prime}e^{\frac{\alpha}{8\pi\gamma l^2_P}E^{\prime}_i m_l}\,$ and peaked both on fluxes $E'_i=8\pi\gamma l_P^2 \bar{j}_i$\footnote{We choose to peak around positive values of $m$'s, such that $m_l=j_l$.} and holonomies $h'=e^{i\bar{\theta}_l m_l}$, where $\bar{j}_i$ and $\bar{\theta}_l$ are labels that, for a given direction $i=x,y,z$ , neither depend on nodes nor on links\footnote{i.e. we keep the labels $l$ only to indicate the direction orthogonal to the surface $S_i$ (pierced by the link $l$) over which the flux $E_i$ is evaluated. Since the direction $l$ depends on $i$, for sake of clarity in (\ref{Nham}) we have dubbed $H'_l$ as $H'_i$.}, i.e. we choose to peak (\ref{semiclassici ridotti inv}) around an homogeneous configuration not restricting to isotropy. Since $E'_i$ and $\bar{\theta}_l$ are local variables, in terms of LQC global physical variables $c_l,p_i$ (for which $\{c_l,p_i\}=8\pi\gamma\delta_{li}$ \cite{Ashtekar:2009um})
we have the following identifications:
\be
E'_i=p_i\,\frac{N_i}{N^3}\,, \qquad \bar{\theta}_l=\frac{c_l}{N_l}\,,\label{lqcvar}
\ee
for a cuboidal graph with $N_i$ number of nodes along one of the three directions $x,y,z$ and $N^3:= N_xN_yN_z\,$. We denote this semiclassical state by $|\Psi_{N_i,H'_i}\rangle\,$ or more explicitely as $\otimes_i |N_i,\bar{j}_i,\bar{\theta}_i \rangle\,.$ Computing the expectation value of the QRLG Hamiltonian operator $\hat{H}$ over these states, in the large $j$ limit we get the following expression \cite{Alesci:2016gub}
\begin{align}
\langle\Psi_{N_i,H'_i}| \hat{H} |\Psi_{N_i,H'_i}\rangle
\approx -\frac{1}{8\pi\gamma^2}\bigg(&N_x\, N_y\,\sqrt{\frac{p_x\;p_y}{p_z}}\;  \sin({\frac{c_x}{N_x}}) \sin({\frac{c_y}{N_y}})
+N_y\, N_z\,\sqrt{\frac{p_y\;p_z}{p_x}}\,\sin({\frac{c_y}{N_y}}) \sin({\frac{c_z}{N_z}})\nonumber\\
&+N_z\, N_x\,\sqrt{\frac{p_z\;p_x}{p_y}}\, \sin({\frac{c_z}{N_z}}) \sin({\frac{c_x}{N_x}})\bigg):=H^{Bianchi}(\{N_i\},\{c_i\},\{p_l\})\label{Nham}
\end{align}
which coincides with the LQC Bianchi I Hamiltonian \cite{Ashtekar:2009vc} as far as we identify the inverse of the number of nodes with the LQC regulator
\be
\mu^{\prime}_i=\frac{1}{N_i}\,. \label{regid}
\ee
It is also worth noticing that this identification is assumed by heuristic arguments in the definition of the improved scheme in LQC, while in QRLG this identification follows naturally within the model due to the presence of the graph structure of the full theory. Therefore, in QRLG we effectively obtain the same semiclassical dynamics as in LQC as soon as we identify the regulator with the inverse number of nodes of the graph at which the states are based. The isotropic limit of (\ref{Nham}) gives the Hamiltonian constraint for the FLRW case:

\be
H^{FLRW}(N^3,c,p):=\braket{\Psi_{N^3,H'}|\hat{H}|\Psi_{N^3,H'}}\approx -\frac{3}{8\pi {\gamma}^2}\sqrt{p} \,{(N^{3})}^{2/3} \sin^2(c {(N^{3})}^{-1/3})\,,\label{FRIEDH1}
\ee
in which we can again define the regulator as 
\be
\mu'=\frac{1}{N}\,. \label{regid2}
\ee
However, the identifications (\ref{regid}) and (\ref{regid2}) lead to the $\mu_{0}$ scheme, since $\hat{H}$ is a non graph-changing Hamiltonian and then the number of nodes is fixed. In the next section we introduce a regularization procedure that at the leading order allows to recover LQC Hamiltonians in $\bar{\mu}$ scheme and provide corrections for the dynamics that we will investigate in the rest of the paper.

Another relevant feature of the QRLG dynamics regards the so called inverse volume corrections that in terms of LQC variables take the following form:
\be
\frac{1}{\sqrt{p_i}}\rightarrow \frac{1}{\sqrt{p_i}}\bigg[1+\frac{N^6}{N^2_{i}}\bigg(\frac{\pi\gamma l_P^{2}}{p_{i}}\bigg)\bigg],
\ee
 which, differently from the LQC ones, are local since depend on the quantum number $j_i$ that enters in $p_i$, i.e. they are invariant under the rescaling of the fiducial cell. The non invariance of the inverse volume corrections in LQC has been criticized by some authors \cite{Bojowald:2008ik} arguing that beyond a minisuperspace approach to cosmology they should be scale invariant, exactly as in QRLG.

\section{Statistical regularization scheme in QRLG}\label{stat}

QRLG-states are represented by cuboidal graphs with $U(1)$ spin labels attached to the links and phases associated to the extrinsic geometry. In this context generic semiclassical states have been constructed by peaking fluxes and holonomies around fixed quantum numbers associated to classical values of the intrinsic and extrinsic geometry in the three directions $x,y,z$. These quantum numbers are\footnote{from now on we write them dropping the overbars.} ($\{N_i\}$,$\{j_i\}$,$\{\theta_i\}$), namely, for a given direction $i$ the total number of nodes associated to a given graph, the spin labels and the argument of the exponential in the classical holonomies (see (\ref{QRLGholo})).

The homogeneous anisotropic sector of the model can be explored from a statistical point of view. Considering as our \textit{microstates} semiclassical coherent states (\ref{semiclassici ridotti inv}) on which homogeneity and anistropy are imposed by peaking them on different set of \textit{microscopic} quantum numbers ($j_{x}$,$j_{y}$,$j_{z}$,$\theta_{x}$,$\theta_{y}$,$\theta_{z}$\,), a \textit{macrostate} is labelled by the set of \textit{collective} variables $(p_x,p_y,p_z;c_x,c_y,c_z)$, related to the microscopic variables by the following (see (\ref{lqcvar})): 
\be
c_{x}=N_{x}\theta_{x}\,, \ \ \ c_{y}=N_{y}\theta_{y}\,, \ \ \ c_{z}=N_{z}\theta_{z}\,,\label{1}
\ee
\be
p_{x}=8\pi\gamma l^{2}_{P}N_{y}N_{z}j_{x}\,, \ \ \ p_{y}=8\pi\gamma l^{2}_{P}N_{x}N_{z}j_{y}\,,  \ \ \ p_{z}=8\pi\gamma l^{2}_{P}N_{x}N_{z}j_{y}, \label{2}
\ee
where $p_{x,y,z}$ and $c_{x,y,z}$ are the total area and the collective $\theta$ along a given direction, respectively. Being free to choose different values of $N_x,N_y,N_z\,,$ to a fixed set of collective variables would correspond several possible combinations of microcopic variables, i.e. the same macrostate corresponds to several different microstates.

The homogeneous and isotropic case can be obtained considering cubical graphs, i.e. graphs with the same number of nodes in the three directions $N_{x}=N_{y}=N_{z}$ and the same peak for the coherent states $j_{x}=j_{y}=j_{z}$, $\theta_{x}=\theta_{y}=\theta_{z}$. In this case the collective variables become ($c,p$), which are related to the microscopic variables by the following:
\be
c=N\theta\,, \label{3}
\ee
\be
p=8\pi\gamma l^{2}_{P}N^{2}j\,. \label{4}
\ee
In \cite{Alesci:2016rmn} has been proposed to assign a weight to all the possible different graphs which represent the same kinematical state defined by a given couple $(p,c)\,$. Here we want to show how this weightning can be seen as a regularization procedure which extend the two previous regularizations in LQC and explore its new resulting physical consequences. 

\subsection{FLRW statistical regularization via volume counting}

Let us start to address the isotropic case. Since $j$ has a minimum value $j_{min}=1/2$, considering two fixed and finite values of $(p,c)$, from \eqref{4} we find that $N$ must have a maximum value $N_{max}$. On the other hand, a non trivial quantum state should have at least one node, namely $N_{min}=1$ which must be associated to a $j_{max}$, thus every $N$ lying in the interval $[N_{min},N_{max}]$ is associated with the couple $(j,\theta)$ such that $(p,c)$ are fixed and provides an indistinguishable semiclassical homogeneous and isotropic state. If we consider a total maximum number of nodes $N^{3}_{max}$ there are several combinations for a given $N^{3}$ which lead to the same state. The total number $\#\Gamma_{N^{3}}$ of indistinguishable graphs for a given $N^{3}$ can be computed by a binomial counting:
\be
\#\Gamma_{N^{3}}=\left(\begin{matrix}N^{3}_{max} \\ N^{3} \end{matrix}\right)\,. \label{bin}
\ee
The normalization factor of this binomial coefficient is easily find to be (for large value of $N_{max}$)
\be
\sum_{N^{3}=1}^{N^{3}_{max}}\left(\begin{matrix}N^{3}_{max} \\ N^{3} \end{matrix}\right)\approx 2^{N^{3}_{max}}\,,
\ee
which tell us that the mean value and the variance of the binomial distribution are respectively:
\be
\langle N^{3}\rangle=\frac{N^{3}_{max}}{2}\,, \ \ \ \sigma^{2}=\frac{N^{3}_{max}}{4}\,.
\ee
Through the assignment of a weight to each configuration, it is possible to define a density matrix whose coefficients give the (classical) probability of the occurence of the various graphs:
\be
\rho^{FLRW}_{N^3}:=\frac{1}{2^{N^{3}_{max}}}\sum_{N^{3}=1}^{N^{3}_{max}}\left(\begin{matrix}N^{3}_{max} \\ N^{3} \end{matrix}\right)|N^{3},j((N^{3})^{2/3},p),\theta((N^{3})^{1/3},c)\rangle\langle N^{3},j((N^{3})^{2/3},p),\theta((N^{3})^{1/3},c)|\,,\label{rhoF}
\ee
where $|N^3,j((N^{3})^{2/3},p),\theta((N^{3})^{1/3},c)\rangle$ is a coherent (micro)state peaked around the classical configuration $(c,p)\,$.
Let us now compute the expectation value of the area operator over this density matrix in the continuum limit, i.e. approximating the binomial with a gaussian with the same mean and variance. Recalling that $\hat{p}$ acts according to (\ref{redei}) we obtain:
\be
\langle \hat{p}\rangle=\frac{Tr(\rho^{FLRW}_{N^3}\,\hat{p})}{Tr{\rho^{FLRW}_{N^3}}}=\frac{8\pi\gamma l_P^{2}\,\int d(N^{3})\,N^{2}\,j(N^2,p)\,e^{-\frac{(N^{3}-N^{3}_{max}/2)^{2}}{N^{3}_{max}/2}}}{\int d(N^{3})\,e^{-\frac{(N^{3}-N^{3}_{max}/2)^{2}}{N^{3}_{max}/2}}}\,. \label{intar}
\ee
To evaluate \eqref{intar} we use the saddle point approximation for large $N^3_{max}$ and keep only the leading order
\be
\langle \hat{p}\rangle\approx 8\pi\gamma l_P^{2}\frac{N^{2}_{max}}{2^{2/3}}j_{0}  \label{arsaddle}
\ee
where $j_0$ is the spin value calculated at the critical point. Identifying $\langle \hat{p}\rangle$ with the $p$ label (\ref{4}), recalling that the existence of $j_{min}$ implies $p=8\pi\gamma l^2_P j_{min}N^2_{max}$, follows that
\be
 8\pi\gamma l_P^{2}\frac{N^{2}_{max}}{2^{2/3}}j_{0}= 8\pi\gamma l_P^{2}N^{2}_{max}j_{min}\label{prelation}
\ee
which implies $2^{2/3}j_{min}=j_{0}$. The (volume counting) statistical regularization consists in taking as an effective hamiltonian the one given by the expectation value of $\hat{H}$ over the density (\ref{rhoF}). Doing this modifies the hamiltonian (\ref{FRIEDH1}) into the following one:
\begin{equation}
H^{FLRW}_{eff,N^3}:=\frac{Tr(\rho^{FLRW}_{N^3}\,\hat{H})}{Tr\rho^{FLRW}_{N^3}}\,,\label{Heffective}
\end{equation}
 whose evaluation at the leading order in the saddle point expansion reads 
 \begin{equation}
 H^{FLRW}_{eff,N^3,0}:=H^{FLRW}(N^3_{max}/2,c,p)= -\frac{3}{8\pi G {\gamma}^2}\sqrt{p} \,{(N_{max}^{3}/2)}^{2/3} \sin^2(c {(N_{max}^{3}/2)}^{-1/3})\,.
 \end{equation} 
 Defining $N^{2}_{max}/2^{2/3}:=1/{\tilde{\mu}}^{2}$ we can cast $H^{FLRW}_{eff,N^3,0}$ in the LQC shape - i.e. like the first term in (\ref{hamlqc}) - and using (\ref{arsaddle}) we find that, up to a constant, the QRLG regulator has the same dependence on $p$ as the $\bar{\mu}$ in LQC:
\be
{\tilde{\mu}}^{2}p=\tilde{\Delta}\,,\label{deltatilde}
\ee
with $\tilde{\Delta}=8\pi\gamma l_P^{2}j_{0}$, i.e. 

\be
\tilde{\Delta}=2^{2/3}\Delta\,.\label{deltaVOL}
\ee

This suggests that the so called LQC area gap should be thought not as a kinematical property but as a property of the particular state under consideration. Further, notice that the derivation in QRLG of this regularization scheme emerges in a natural way without importing structures from the full theory.

\subsection{$\mu_0$ and $\bar{\mu}$ scheme in QRLG}

We are in a position to show how the LQC schemes emerge from QRLG, which are the differences with the statistical regularization and why the latter can be seen as an extension of the former. From a QRLG point of view, all the regularization schemes can be thought as coming from a relation between the area $\langle \hat{p}\rangle$ and the microscopical quantum numbers $(N^{2},j)$ through a selected density matrix whose coefficients represent the probability of having the different allowed combinations of quantum numbers $(N,j)$ such that the global variables $(p,c)$ take the same values. In this way we can easily build the two density matrices associated respectively to the $\mu_{0}$ and $\bar{\mu}$ scheme:
\be
\rho^{FLRW}_{\mu_{0}}:=\delta_{N^{3},N^{3}_{min}}|N^{3},j((N^{3})^{2/3},p),\theta((N^{3})^{1/3},c)\rangle\langle N^{3},j((N^{3})^{2/3},p),\theta((N^{3})^{1/3},c)|\,, \label{denmu0}
\ee
\be
\rho^{FLRW}_{\bar{\mu}}:=\delta_{N^{3},N^{3}_{max}}|N^{3},j((N^{3})^{2/3},p),\theta((N^{3})^{1/3},c)\rangle\langle N^{3},j((N^{3})^{2/3},p),\theta((N^{3})^{1/3},c)|\,, \label{denmubar}
\ee
whose continuum limit is simply defined substituting the Kronecker delta with the Dirac delta. Looking at expressions (\ref{Nham}), (\ref{regid}) and (\ref{FRIEDH1}) is easy to realize that \textit{a} $\mu_0$ scheme\footnote{indeed the $\mu_0$ obtained here has a different numerical value compared to the one in LQC, which is $3\sqrt{3}$ (see (\ref{3rad3})).} in QRLG is obtained, indeed, computing expectation values over the state (\ref{denmu0}) and recalling $N_{min}=1\,$ we get $\mu^{\prime}=1:=\mu_0\,.$

The $\rho^{FLRW}_{\bar{\mu}}$ instead selects more naturally a privileged homogeneous and isotropic maximally coarse-grained state. The expectation value of the area operator over this density matrix reads
\be
\langle \hat{p}\rangle=\frac{Tr(\rho^{FLRW}_{\bar{\mu}}\,\hat{p})}{Tr{\rho^{FLRW}_{\bar{\mu}}}}=8\pi\gamma l_P^{2}\,N^{2}_{max}\,j_{min}
\ee
and a $\bar{\mu}$ scheme is recovered, since setting $\bar{\mu}:=1/N_{max}$, it holds $\bar{\mu}^{2}p=\Delta$. 

In closing, we have seen the LQC schemes select a privileged state neglecting the contribution of all the others which give a indistinguishable global geometry, i.e. neglecting the quantum fluctuations associated to $N\,$. QRLG instead selects the most probable state assigning a statistical weight to all the possible indistinguishable configurations, in this way the state which statistically has the best properties of homogeneity and isotropy is the one centered in $N^{3}_{max}/2$. However, computing the expectation value of the various operators over the QRLG density matrix it is possible to go over the first order approximation and take into account the fluctuations due to the less probable homogeneous and isotropic states. Since the two privileged states selected by the LQC regularization schemes turn out to be in the two opposite tail of the resulting (finite) gaussian, they can be seen just as particular choices of the statistical regularization in QRLG.

\subsection{The emergent-bouncing universe}

The physical consequences of the statistical regularization were analysed for the first time in \cite{Alesci:2016xqa} where the ratio (\ref{Heffective}) was evaluated at the next to the leading order of the saddle point expansion. The result gives the following Hamiltonian

\be
H^{FLRW}_{eff,N^3,1}=-\frac{3}{8\pi{\gamma}^2}\sqrt{p}\left(\frac{N^2_{max}}{2^{2/3}}\sin^{2}(\frac{2^{1/3}c}{N_{max}}) + \frac{c^{2}}{9N^3_{max}}\cos(2\frac{2^{1/3}c}{N_{max}})-\frac{2^{1/3}}{18 N_{max}}\sin^2(\frac{2^{1/3}c}{N_{max}})\right)\,. \label{hqrlgNpc}
\ee
To study the resulting dynamics it is convenient to switch to a new pair of canonical variables related to $(p,c)$ in the following way:
\be
v=\frac{p^{3/2}}{2\pi\gamma}
\,, \ \ \ b=\frac{c}{p^{1/2}}\,,\label{bev}
\ee
anf the only non vanishing Poisson brackets read:
\be
\{v,b\}=-2\,.
\ee
Writing (\ref{hqrlgNpc}) in terms of these new variables, using (\ref{deltatilde}) and adding the usual massless scalar field coupling, we get 
\be
H^{FLRW}_{eff,N^3,1}+H_{\phi}=-\frac{3v}{4\tilde{\Delta}\gamma}\,\sin^2(\sqrt{\tilde{\Delta}}b)+\frac{P_{\phi}^2}{4\pi\gamma v} - \frac{ b^2{\tilde{\Delta}}^{3/2}}{48\pi\gamma^2}\cos(2\sqrt{\tilde{\Delta}}b)+\frac{\sqrt{\tilde{\Delta}}}{48\pi\gamma^2 }\sin^2(\sqrt{\tilde{\Delta}} b)\,,\label{ciao2}
\ee
computing the associated equations of motion and neglecting terms that are subdominant in a $1/v$ expansion\footnote{In writing eq. (\ref{F5}) the last term in the rhs of (\ref{ciao2}) has been neglected, too. The presence of that term amounts just in shifting a bit the coefficient in front of the LQC-like term, i.e. in the first term in the rhs, and it does not affect the dynamics \cite{Theguys}. Anyway, the numerical study in section \ref{result} has been done keeping also that contribution.} we find the following modified Friedmann equation
\be
\Bigg(\frac{\dot{v}}{3v}\Bigg)^{2}=\left(\frac{8\pi}{3}\\\rho_{\rm m}+\frac{\rho_{\rm g}}{\gamma^2}\right)(1-2\Omega_{\rm g})^{-1}\left(1-\frac{\Omega_{\rm m}-\Omega_{\rm g}}{1-2\Omega_{\rm g}}\right)\,,\label{F5}
\ee
where the new quantities are defined as follows:
\begin{align}
&\rho_{\rm g}:=-\frac{b^2{\tilde{\Delta}}^{3/2}}{18 V}\,,\ \  \bar{\rho}_{\rm cr}:=-\frac{1}{\tilde{\Delta}}\,,\ \ \ \ \ \ \rho_{ m}:=\frac{P_{\phi}^{2}}{2V^{2}}\,,\ \rho_{cr}:=\frac{3}{8\pi\gamma^{2}\tilde{\Delta}}\,, \ \  
&\Omega_{\rm g}:=\frac{\rho_{\rm g}}{\bar{\rho}_{\rm cr}}\, \ \ \ \Omega_{\rm m}:=\frac{\rho_{\rm m}}{\rho_{\rm cr}}\,,\label{V1}
\end{align}
where $V$ is the physical volume, $V=2\pi\gamma v\,$. The density $\rho_{\rm g}$ is directly related to the cosine term in the Hamiltonian and it can be interpreted as a geometric negative energy density whose origin is purely quantum gravitational, $\bar{\rho}_{cr}$ is the critical energy density at which a universe without matter would undergo a bounce. The former is a pure quantum gravity correction since it depends on the ratio $l_P^{3}/V$, hence it is totally negligible for large universes, but it deeply affects the dynamics in the quantum regime. Differently from LQC there are two conditions that lead to a stationary point:
\be
\Omega_{\rm g}+\Omega_{\rm m}=1\,, \ \ \ \Omega_{\rm g}=\Omega_{\rm m}\,, \label{cond}
\ee
the first is the same bounce condition of LQC: when the sum of the ratio between $\rho_{\rm g}/\bar{\rho}_{cr}$ and $\rho_{ m}/\rho_{cr}$ is equal to $1$, the universe bounces, hence the system reaches a global minimum. However a new possibility is allowed when quantum gravity effects compensate the evolution driven by the matter content, this is encoded in the second condition of \eqref{cond}. In this regime the evolution can reach even maxima and the LQC pre-bounce dynamics is replaced by consecutive oscillations with decreasing local maxima as we go back in time (see top panel of fig.\ref{vplot} in section \ref{result}). 

We stress this emergent-bouncing phase is not postulated at all but it naturally emerges from a fundamental quantum gravity model for the FLRW universe. Furthermore it sheds light for the first time over a possible dynamical quantum effect of the gravitational field near the Planck scale: it starts to act as a negative energy density behaving as a ``repulsive force''. 

The modification of the LQC bouncing scenario is not the only physical consequence of the statistical regularization, indeed the cosine term in the QRLG Hamiltonian breaks a classical symmetry still present in the $\bar{\mu}$ LQC Hamiltonian (only if one neglects the inverse volume corrections): the freedom on the rescaling of the fiducial cell. However, as we will discuss in detail, this fact is unavoidable for the appearence of a new physical scale, namely the Planck scale. Still, within the QRLG dynamics this symmetry is broken only when the size of the universe approaches the Planck volume and as the universe becomes classical, homogeneity is perfectly restored since this term is negligible and we can factor out $V_0$ recalling $P_{\phi}=V\dot{\phi}\,.$

Let us point out the differences about the rescaling symetry breaking between the LQC $\mu_{0}$ scheme and QRLG. In LQC the cell dependence had an influence on physical quantities like $\rho_{cr}$ since it had a dependence on $1/P_{\phi}$, hence for different values of the fiducial cell, the density at which the bounce occurs could be in principle shifted to any values. This problem was impossible to solve even considering a compact topology since taking realistic initial cosmological data the value of the density predicted for the bounce was extremely classical. In QRLG the term responsible for the emergent-bouncing phase depends on the ratio $V_{P}/V$, and is telling us that the theory must take into account dynamical quantum gravitational effects close to the Planck scale when $V\rightarrow V_P\,$. Furthermore, differently from the LQC $\mu_{0}$ scheme, if we consider a compact topology no more interpretative problems arise since the density at which the bounce occurs is always fixed. If we want to consider a non compact topology like $\mathbb{R}^{3}$ it is true that from a matematical point of view it is not possible to fix univoquely the cell $V_{0}$ since in an open homogeneous spacetime at least at the classical level different regions of the universe should have an indistinguishable evolution. In this regard the emergent-bouncing universe is in full agreement with this picture since for classical universe the ratio $V_{P}/V$ is completely negligeable and the rescaling symmetry is completely restored. If we allow a breaking of classical homogeneity at Planck scale, the emergent-bouncing universe doesn't present any conceptual issue, since the sistem generated by (\ref{ciao2}) is completely determined once initial conditions for $(b,v)$ are given and then is not the value of $V_0$ itself to be physically relevant but only the physical volume $V$. One could argue that in an open topology the total volume of the universe should tend to infinity hence the effect of the new term is negligible. In our opinion this claim is not physically relevant since even for an infinite universe only a finite region is causally connected to us. Different initial data describe inequivalent universes with different global physical volumes $V$ and a physically motivated choice of initial data should be related to the biggest region of the universe causally connected to us that leads to predictions compatible with observations. Dynamics generated by (\ref{ciao2}) is predictive and at most requires to experimentally fix the initial value of $V$.

\section{Bianchi $I$ statistical regularization in QRLG}\label{Bianchi}

We are now ready to apply the QRLG statistical regularization scheme to a non isotropic case, to recover the LQC improved dynamics of Bianchi $I$ and its new corrections. Let us start by recalling the effective LQC Bianchi $I$ Hamiltonian:
\be
H_{LQC}^{Bianchi}\sim \sqrt{\frac{p_xp_y}{p_z}}\frac{\sin({\bar{\mu}_{x}c_x})\sin(\bar{\mu}_y c_y)}{\bar{\mu}_x\bar{\mu}_y}+\sqrt{\frac{p_xp_z}{p_y}}\frac{\sin(\bar{\mu}_{x}c_x)\sin(\bar{\mu}_z c_z)}{\bar{\mu}_x\bar{\mu}_z}+\sqrt{\frac{p_zp_y}{p_x}}\frac{\sin(\bar{\mu}_{z}c_z)\sin(\bar{\mu}_y c_y)}{\bar{\mu}_z\bar{\mu}_y}\,. \label{hamblqc}
\ee

From the QRLG perspective, as in the FLRW case, the $\bar{\mu}$ LQC regularization scheme is associated to a particular choice of a kinematical state which from a semiclassical point of view corresponds only to a particular microstate.  To apply the statistical regularization scheme we have to classify microstates looking at the quantum numbers that caracterize a Bianchi $I$ semiclassical state using relations (\ref{1}) and (\ref{2}). This allows to build the density matrix $\rho^{Bianchi}_{\bar{\mu}}\,$, extention of (\ref{rhoF}), suitable to describe the Bianchi $I$ case.

From (\ref{1}) and (\ref{2}) we see that each spin $j_i$ introduces a bound on the product $N_j N_k\,(i\neq j\neq k)$. This suggests to consider a new set of indipendent variables

\begin{equation}
A_x:= N_y N_z\,,\quad A_y:=N_z N_x\,,\quad A_z:= N_x N_y\,,
\end{equation}
in terms of which (\ref{1}) and (\ref{2}) read as follows:
\begin{equation}
p_i=8\pi\gamma l^2_P\, A_i j_i\,,\quad c_i=\sqrt{\frac{A_jA_k}{A_i}} \theta_i \qquad \mbox{\,where\,} i\neq j\neq k\,, \label{kspa}
\end{equation}
then the quantum numbers which characterize the microstates lie in the intervals contained in the three sets $\{[A_i^{min},A_i^{max}],[j_i^{min},j_i^{max}]\}$.
If we want to recover the improved scheme of LQC, we have to consider the most coarse-grained kinematical configuration associated to a particular state corrisponding to the following density matrix:
\be
\rho^{Bianchi}_{\bar{\mu}}:=\prod_i \delta_{A_i A_i^{max}}\,|A_i,j_i(A_i,p_i),\theta_i(\{A_i\},c_i)\rangle \langle A_i,j_i(A_i,p_i),\theta_i(\{A_i\},c_i)|\,,\label{dmatbq}
\ee
the expectation value of $\hat{H}$ over the state \eqref{dmatbq} gives the QRLG Bianchi $I$ Hamiltonian
\begin{eqnarray}
H^{Bianchi}(\{A_i^{max}\},\{c_i\},\{p_i\})&\equiv & -\frac{1}{8\pi\gamma^2}\left[\right.A^{max}_z \sqrt{\frac{p_x p_y}{p_z}}\sin(c_x\sqrt{\frac{A^{max}_x}{A^{max}_y A^{max}_z}})\sin(c_y\sqrt{\frac{A^{max}_y}{A^{max}_x A^{max}_z}})+ \nonumber \\
&+&A^{max}_x \sqrt{\frac{p_y p_z}{p_x}}\sin(c_y\sqrt{\frac{A^{max}_y}{A^{max}_x A^{max}_z}})\sin(c_z\sqrt{\frac{A^{max}_z}{A^{max}_x A^{max}_y}}) 
+\nonumber \\
&+&A^{max}_y \sqrt{\frac{p_z p_x}{p_y}}\sin(c_z\sqrt{\frac{A^{max}_z}{A^{max}_x A^{max}_y}})\sin(c_x\sqrt{\frac{A^{max}_x}{A^{max}_y A^{max}_z}})\left.\right]\,.\label{hambqrlg}
\end{eqnarray}
Let us consider the set of the collective labels $(p_x,p_y,p_z)$ for a state associated to $(A_x^{max},A_y^{max},A_z^{max})$, they read:
\be
p_i=8\pi \gamma l^2_P\, A_i^{max} j_i^{min}\label{67}
\ee
where the three independent variables $\{A_i^{max}\}$ can be rewritten in terms of the number of nodes as
\be
A_x^{max}=N_y^{max}N_z^{max}\,, \ \ \ A_y^{max}=N_z^{max}N_x^{max}\,, \ \ \ A_z^{max}=N_x^{max}N_y^{max}\,,
\ee
then identifying the LQC regulator with the inverse of the number of nodes
\be
\mu_i=\frac{1}{N_i^{max}}
\ee
and using $j_x^{min}=j_y^{min}=j_z^{min}=1/2$, from (\ref{67}) we get the following relations:
\be
\mu_{y}\mu_{z}p_{x}=\Delta\,, \ \ \ \mu_{y}\mu_{x}p_{z}=\Delta\,, \ \ \ \mu_{x}\mu_{z}p_{y}=\Delta\,, \label{mubarbq}
\ee
where $\Delta$ is the minimum eigenvalue of the Area operator in QRLG. The relations \eqref{mubarbq} give exactly the improved version of the regulators for the Bianchi $I$ universe in LQC \cite{Ashtekar:2011ni,Ashtekar:2009vc}. We stress that this derivation from QRLG doesn't require any input from the full theory but it is self-consistent and is based on the only assumption that the most coarse-grained  configuration is the privileged one. Anyway, as for the FLRW case, the $\bar{\mu}$ scheme ignores several semiclassical indistinguishable microstates that lead to the same geometrical configuration. Indeed, the use of the Kronecker delta forces the number of nodes to be a classical number, on the contrary, if we consider a non trivial density matrix we can take into account the quantum nature of $N_{i}$, namely we can deal with the spread associated to a chosen distribution. We do this now.

The extention of the density matrix (\ref{rhoF}) to the Bianchi $I$ case requires to weight the occurrence of the microstates using a binomial counting, namely to use the following density matrix
\begin{equation}
\rho^{Bianchi}:=\prod_i \sum_{A_i=1}^{A_i^{max}}\left(\begin{matrix} A_i^{max}\\ A_i \end{matrix}\right)|A_i,j_i(A_i,p_i),\theta_i(\{A_i\},c_i)\rangle \langle A_i,j_i(A_i,p_i),\theta_i(\{A_i\},c_i)|\,. \label{dbmai}
\end{equation}

In the continuum limit and at the leading order of the saddle point expansion, the regularization of the QRLG Hamiltonian (\ref{hambqrlg}) with this density matrix gives  $H^{Bianchi}(\{A_i^{max}/2\})$, i.e the hamiltonian evaluated at the most probable value of the three gaussians. Analogously to (\ref{arsaddle}), it follows that
\be
\langle \hat{p}_i \rangle= \frac{Tr(\rho^{Bianchi} \hat{p}_i)}{Tr \rho^{Bianchi}} =8\pi\gamma l_P^2\, \frac{A_i^{max}}{2} j_i^{0}\,
\ee
the identification of $\langle \hat{p}_i\rangle$ with the labels $p_i$ (\ref{67}) and the evaluation of the Hamiltonian 

\begin{equation}
H^{Bianchi}_{eff}:=\frac{Tr(\rho^{Bianchi}\,\hat{H})}{Tr\rho^{Bianchi}}
\label{HBianchieff}
\end{equation} 
allow to cast its leading order $H^{Bianchi}_{eff,0}:=H^{Bianchi}(\{A_i^{max}/2\})$ in the LQC form (\ref{hamblqc}) once we set 
\be
\mu'_{i}:=\frac{\sqrt{2}}{{N_i^{max}}}\,,
\ee
that implies
\be
\mu'_{y}\mu'_{z}p_{x}=\Delta'\,, \ \ \ \mu'_{y}\mu'_{x}p_{z}=\Delta'\,, \ \ \ \mu'_{x}\mu'_{z}p_{y}=\Delta'\,, \label{mubarbqrlg}
\ee
where 
\be
\Delta'=2\Delta\,.\label{deltaAREE}
\ee
 The relations (\ref{mubarbqrlg}) have the same functional form of the LQC ones even if the value of $\Delta'$ is different, since the statistical counting does not assign the most probable value to the most refined state. The use of a non trivial density matrix introduces a spread for $N_i$ that now must be taken into account  evaluating the saddle point expansion of the Hamiltonian. This correction gives a departure from LQC by means of a new effective hamiltonian for Bianchi $I$ that coincides at the leading order with the one of LQC, as we are ready to show.

Considering the continuous approximation and performing the usual substitutions 
\be
A_x:=A_x^{max} S_{x}\,,\quad A_y:=A_y^{max} S_{y}\,,\quad A_z:=A_z^{max} S_{z}\,,\quad \lambda:=A_x^{max}A_y^{max}A_z^{max}\,,\label{substitutions}
\ee
we get
\begin{equation}
H^{Bianchi}_{eff}(\lambda)\approx\frac{\lambda\int dS_x dS_y dS_z\,H^{Bianchi}\, e^{\lambda\left[-\frac{2(S_x-1/2)^2}{A_y^{max}A_z^{max}}-\frac{2(S_y-1/2)^2}{A_z^{max}A_x^{max}}-\frac{2(S_z-1/2)^2}{A_x^{max}A_y^{max}}\right]}}{Tr \rho^{Bianchi}(\lambda)} \label{intBianchi}
\end{equation}
which it is now in the form that allows us to apply the multidimensional saddle point formula and to write
\begin{equation}
H^{Bianchi}_{eff}(\lambda)\approx \frac{\lambda\,\left(\frac{2\pi}{\lambda}\right)^{3/2}\,\frac{e^{\lambda \phi(crit)}}{\sqrt{|det\phi''(crit)|}}\left[H^{Bianchi}(crit)-\frac{1}{2\lambda}H''^{Bianchi}_{ij}(crit)\phi''^{-1}_{ij}(crit)+... \right]}{Tr\rho^{Bianchi}(\lambda)}\label{saddle3d}
\end{equation}

where $$\phi''^{-1}_{ij}(crit)=-1/4\, diag \left(A_y^{max}A_z^{max},\,A_x^{max}A_z^{max},\,A_y^{max}A_x^{max}\right)\,,\quad crit:=(1/2,1/2,1/2)$$ and we end with the result
\begin{equation}
\begin{split}
&H^{Bianchi}_{eff,1}=H^{Bianchi}_{eff,0}+\nonumber\\
&\frac{1}{192\sqrt{2}\pi\gamma^2\lambda^2}\left[ \right. \,A_z^{max}\sqrt{\lambda}\, c_y\cos(c_y\sqrt{\frac{2A_z^{max}}{A_x^{max}A_y^{max}}})\left[\right. \alpha_1 \sqrt{\frac{p_x p_y}{p_z}} \sin(c_x\sqrt{\frac{2A_y^{max}}{A_x^{max}A_z^{max}}})\,+\,\alpha_2 \sqrt{\frac{p_y p_z}{p_x}}\sin(c_z\sqrt{\frac{2A_x^{max}}{A_y^{max}A_z^{max}}})\left.\right] \nonumber \\
&-\,(A_x^{max})^2\,\sqrt{\frac{A_z^{max}A_y^{max}}{A_x^{max}}} c_z \cos(c_z\sqrt{\frac{2A_x^{max}}{A_y^{max}A_z^{max}}}) \left[\right.\left[\right.\alpha_3 \sqrt{\frac{p_x p_z}{p_y}} \sin(c_x\sqrt{\frac{2A_y^{max}}{A_x^{max}A_z^{max}}})\,+\,\alpha_4 \sqrt{\frac{p_y p_z}{p_x}} \sin(c_y\sqrt{\frac{2A_z^{max}}{A_x^{max}A_z^{max}}})\left.\right] \nonumber\\
&+\,(A_y^{max})^2\,\sqrt{\frac{A_x^{max}}{A_y^{max}A_z^{max}}} c_x \cos(c_x\sqrt{\frac{2A_y^{max}}{A_x^{max}A_z^{max}}})\left[\right.\alpha_5 \sqrt{\frac{p_x p_y}{p_z}} \sin(c_y\sqrt{\frac{2A_z^{max}}{A_x^{max}A_y^{max}}})\,+\,\alpha_6 \sqrt{\frac{p_x p_z}{p_y}} \sin(c_z\sqrt{\frac{2A_x^{max}}{A_z^{max}A_y^{max}}}) \left.\left.\right]\left.\right]\,\right]\nonumber\\ 
\end{split}\label{bordello}
\end{equation}
where
\begin{eqnarray}
&\alpha_1 &:=A_x^{max}[A_x^{max}(A_y^{max}-3A_z^{max})\,+\,A_y^{max}A_z^{max}]\,,\quad \alpha_2:=A_y^{max}[-3A_y^{max}A_z^{max}+A_x^{max}(A_y^{max}+A_z^{max})]\nonumber\\
&\alpha_3 &:=-A_z^{max}[A_x^{max}(A_y^{max}-3A_z^{max})+A_y^{max}A_z^{max}]\,,\quad \alpha_4:=A_y^{max}[3A_x^{max}A_y^{max}-A_x^{max}A_z^{max}-A_y^{max}A_z^{max}]\nonumber\\
&\alpha_5 &:= -A_x^{max}[3A_x^{max}A_y^{max}-A_x^{max}A_z^{max}-A_y^{max}A_z^{max}]\,,\quad \alpha_6:=A_z^{max}[-3A_y^{max}A_z^{max}+A_x^{max}(A_y^{max}+A_z^{max})]\,.\nonumber\\
\end{eqnarray}
To keep space, in the latter we have written only terms of order $\lambda^{-1/6}$ , i.e the leading terms among all\footnote{Indeed, because of substitutions (\ref{substitutions}), notice the Bianchi hamiltonian and the hessian matrix keep terms proportional to $\lambda^{0}\,, \lambda^{1/6}$ and $\lambda^{2/3}$ respectively.} the next to the leading ones. The detailed study of the dynamics generated by $H^{Bianchi}_{eff,1}$ is beyond the scope of this work and will be addressed in a future paper under preparation.

\section{FLRW statistical regularization via area counting}\label{AREAFRID}

The statistical treatment of the Bianchi $I$ case required a counting based on coordinate areas with number of nodes $A_i$ contained in the cuboidal lattice. The original computation (\ref{Heffective}) leading to the emergent bouncing universe was instead based on counting coordinate volumes contained in a cubical lattice with $N^3$ nodes. The counting choice, at this level, is still arbitrary since only the solution of the hamiltonian constraint can restrict the freedom in choosing the density matrix $\rho$. Within this freedom one may wonder if the emergent bouncing scenario survives where different countings are considered. 

In this section we compute the effective FLRW hamiltonian as it results by the statistical regularization defined by the ``area counting'' suggested by the anisotropic case. To begin with, lets rewrite here the Hamiltonian (\ref{FRIEDH1}) expliciting areas rather than volumes, i.e. 
\be
H^{FLRW}=-\frac{3}{8\pi  {\gamma}^2}\sqrt{p}\, A \sin^2(A^{-1/2}c)
\ee
where $A\equiv N^2\,$. Using the following density matrix
\begin{equation}
\rho^{FLRW}_{A}:= \frac{1}{2^{A^{max}}}\sum_{A=1}^{A^{max}} \left(\begin{matrix} A^{max} \\ A \end{matrix} \right)|A,j(A,p),\theta(A,c) \rangle \langle A,j(A,p),\theta(A,c) |\,,
\end{equation}
we are now interested in computing the following quantity
\be
H^{FLRW}_{eff,A}:=\frac{Tr(\rho^{FLRW}_{A} \hat{H})}{Tr\rho^{FLRW}_A}\equiv \frac{-\frac{3}{8\pi {\gamma}^2}\frac{\sqrt{p}}{2^{A^{max}}}\sum_{A=1}^{A^{max}} \left(\begin{matrix} A^{max} \\ A \end{matrix}\right) \, A \sin^2(A^{-1/2}c) }{Tr\rho^{FLRW}_A}\,.\label{HeffFriedAREE}
\ee
The contribution to (\ref{HeffFriedAREE}) at the next to the leading order in the saddle point approximation for large $A$ is easily found to be
\begin{equation}
H^{FLRW}_{eff,A,1}=-\frac{3}{8\pi  {\gamma}^2}\sqrt{p}\,\left[ \frac{ \sin^2( \frac{c\sqrt{2}}{ N_{max} })} {(\frac{\sqrt{2}}{N_{max}})^2} - \frac{1}{8 N^2_{max}} \left(\frac{cN_{max}}{\sqrt{2}} \sin(2 \frac{c\sqrt{2}}{ N_{max} }) - 2 c^2 \cos(2\frac{c\sqrt{2}}{N_{max}})\right)\right]\,,\label{HFeffAREE1}
\end{equation}
in which we have restored $A^{max}=N^2_{max}$ in order to make clear that the hamiltonian (\ref{HFeffAREE1}) differs, up to the leading order, from what we found counting the variable $N^3$, i.e. from hamiltonian (\ref{hqrlgNpc}).

Finally, lets write (\ref{HFeffAREE1}) in terms of the variables $v$ and $b$ (see (\ref{bev})) together with the contribution of a massless scalar field $\phi\,$,

\be
H^{FLRW}_{eff,A,1}(v,b)+H_{\phi}=-\frac{3v}{4\gamma\Delta'}\sin^{2}(\sqrt{\Delta'}b)+\frac{P_{\phi}^{2}}{4\pi\gamma v}+\frac{3\sqrt{\Delta'}b(2\pi\gamma v)^{1/3}}{128\pi\gamma^{2}}\sin(2\sqrt{\Delta'}b)-\frac{3(2\pi\gamma v)^{1/3}\Delta' b^{2}}{64\pi\gamma^{2}}\cos(2\sqrt{\Delta'}b)\,,\label{HFriedareebev}
\ee
from which we get the following eqq. of motion:

\be
\begin{split}
\dot{v}=\frac{3v}{2\gamma\sqrt{\Delta'}}\sin(2\sqrt{\Delta'}b)-\frac{3\sqrt{\Delta'}(2\pi\gamma v)^{1/3}}{64\pi\gamma^2}\sin(2\sqrt{\Delta'}b)&+\frac{3(2\pi\gamma v)^{1/3}\Delta' b}{32\pi\gamma^{2}}\cos(2\sqrt{\Delta'} b)\,+\nonumber\\
&-\frac{3(2\pi\gamma v)^{1/3}\Delta'^{3/2}b^{2}}{16\pi\gamma^{2}}\sin(2\sqrt{\Delta'}b)\,;
\end{split}
\ee

\be
\dot{b}=-\frac{3}{2\gamma\Delta'}\sin^{2}(\sqrt{\Delta'}b)-\frac{P_{\phi}^{2}}{2\pi\gamma v^{2}}-\frac{(2\pi\gamma )^{1/3}\sqrt{\Delta'}b}{64\pi\gamma^{2}v^{2/3}}\sin(2\sqrt{\Delta'} b)+\frac{(2\pi\gamma )^{1/3}\Delta' b^{2}}{32\pi\gamma^{2}v^{2/3}}\cos(2\sqrt{\Delta'}b)\,.
\ee
The associated dynamics is studied in section \ref{result}.

\section{FLRW from the Bianchi $I$ reduced density matrix}\label{REDUCED}

We show here how from states \eqref{dmatbq} and \eqref{dbmai} describing Bianchi $I$ universe respectively in LQC and in QRLG, we can get the (reduced) density matrices for the Friedmann case. Let us consider the density matrix \eqref{dmatbq} and the associated reduced density matrix obtained tracing it over two of the three independent variables (e.g. over $A_y,A_z$):
\be
\rho_{red}=\delta_{A_x A_x^{max}}|A_x,j_x(A_x,p_x),\theta_x(\{A_i\},c_x)\rangle \langle A_x,j_x(A_x,p_x),\theta_x(\{A_i\},c_x)|\,. \label{redmat}
\ee
The expectation value of the QRLG Bianchi $I$ Hamiltonian (\ref{hambqrlg}) over the state $\rho_{red}$ reads :
\be
\begin{split}
Tr(\rho_{red}\,&H^{Bianchi})=\nonumber\\
&-\frac{\delta_{A_x A_x^{max}}}{8\pi\gamma^2}\left[\right. A^{}_z \sqrt{\frac{p_x p_y}{p_z}}\sin(c_x\sqrt{\frac{A^{}_x}{A^{}_y A^{}_z}})\sin(c_y\sqrt{\frac{A^{}_y}{A^{}_x A^{}_z}})
+A^{}_x \sqrt{\frac{p_y p_z}{p_x}}\sin(c_y\sqrt{\frac{A^{}_y}{A^{}_x A^{}_z}})\sin(c_z\sqrt{\frac{A^{}_z}{A^{}_x A^{}_y}}) +
\nonumber \\
&+A^{}_y \sqrt{\frac{p_z p_x}{p_y}}\sin(c_z\sqrt{\frac{A^{}_z}{A^{}_x A^{}_y}})\sin(c_x\sqrt{\frac{A^{}_x}{A^{}_y A^{}_z}})\left.\right]\label{dmai}
\end{split}
\ee
and in the isotropic limit $c_x=c_y=c_z\,$, $p_x=p_y=p_z\,$, $A_x=A_y=A_z$ gives:
\be
-\delta_{A_x A_x^{max}}\,\frac{3}{8\pi\gamma^2}\sqrt{p}\,A \sin^2(A^{-1/2}c)\,,
\ee
which is exactly the QRLG hamiltonian (\ref{FRIEDH1}) and from which it is possible to recover the LQC improved scheme. 

Analogous result holds starting for the Bianchi $I$ effective hamiltonian defined by means of the statistical regularization. Let us consider the density matrix \eqref{dbmai}, if we trace over two of the three independent variables we get:
\be
\rho_{red}^{stat}=\sum_{A=1}^{A_x^{max}} \left(\begin{matrix} A_x^{max} \\ A \end{matrix}\right)|A_x,j_x(A_x,p_x),\theta_x(\{A_i\},c_x)\rangle \langle A_x,j_x(A_x,p_x),\theta_x(\{A_i\},c_x)|\,,
\ee
using this density matrix for evaluating the expectation value of the QRLG Bianchi $I$ Hamiltonian and taking the isotropic limit we find:
\be
-\frac{3}{8\pi\gamma^2}\frac{\sqrt{p}}{{2^{A^{max}}}}\sum_{A=1}^{A^{max}} \left(\begin{matrix} A^{max} \\ A \end{matrix}\right)A \sin^2(A^{-1/2}c)
\ee
which coincides with the FLRW effective Hamiltonian (\ref{HeffFriedAREE}). This last result suggested the use of the new counting method based on areas $\{A_i\}$. This counting is different from the first proposed in \cite{Alesci:2016rmn} indeed the latter was based on counting the number of nodes $N^{3}$ cointained in the total cubical volume. If for FLRW this counting is perfectly fine for classifying all the possible microstates, an extention for Bianchi $I$ turns out to be impossible since one cannot recover all the kinematical space from a single volume variable of a three dimensional rectangoloid without knowing the areas of its faces $N_iN_j$ and the lenghts of its sides $N_i$. Instead, counting areas the extention to Bianchi $I$ is straightforward. Moreover, the density matrix associated to the new counting can be reduced to the one used for FLRW in the isotropic limit, supporting the generality of this nex counting.

Finally, a natural question about the different dynamics associated to the Friedmann Hamiltonian (\ref{ciao2}) \cite{Alesci:2016xqa} and the new one obtained counting the areas arises. In the following section we show how in both cases dynamics near the LQC-bounce is qualitatively the same, a result that supports the generality of the emergent-bouncing universe in QRLG.

\section{FLRW effective dynamics: numerical study}\label{result}

Here we show the result of the numerical evolution for the dynamics associated to the QRLG effective Hamiltonians (\ref{ciao2}), (\ref{HFriedareebev}) and the standard LQC one for the FLRW case.
\begin{figure}[H]
\centering
\includegraphics[width=18cm]{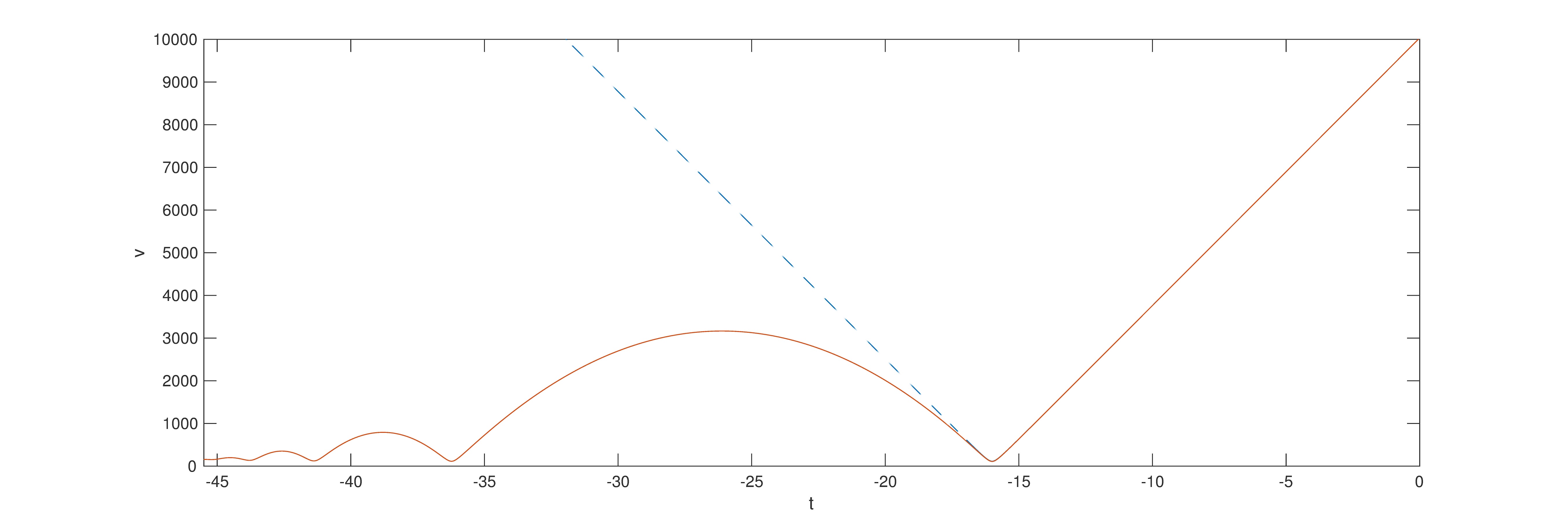} 
\includegraphics[width=18cm]{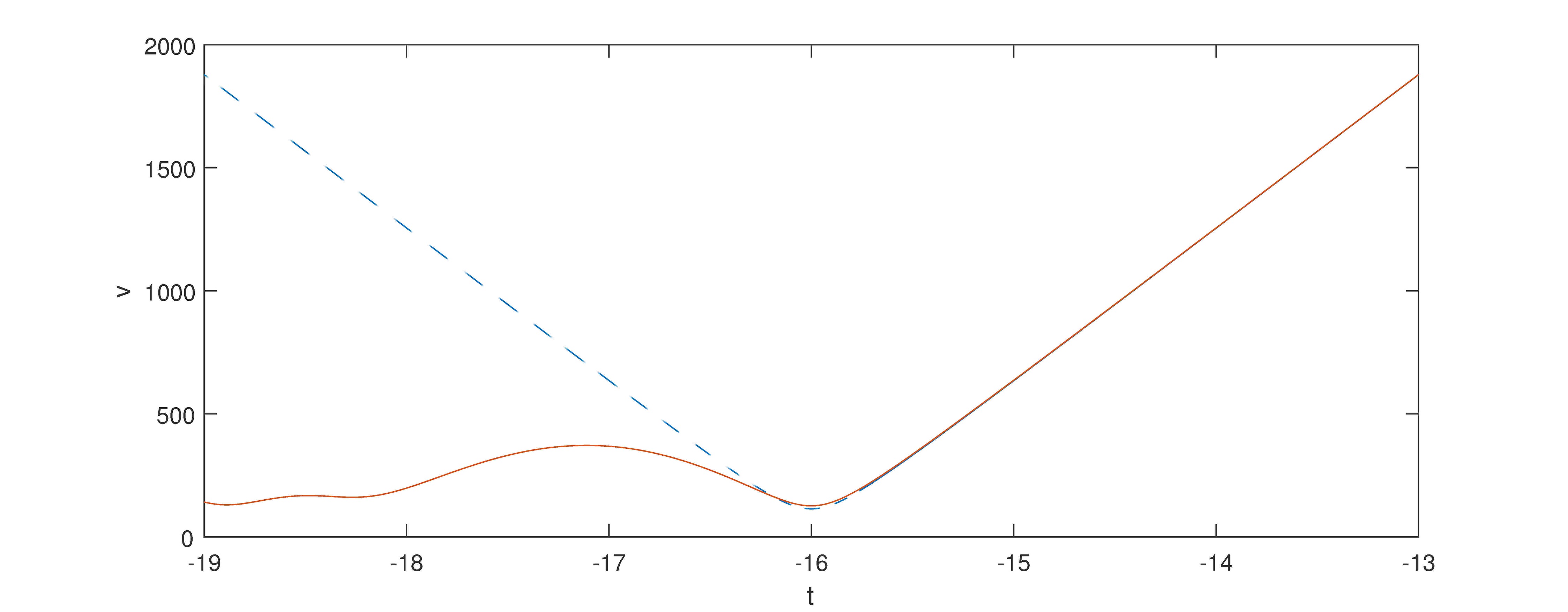}
\caption{From top to bottom the evolution of $v$ vs t according to the QRLG effective Hamiltonians regularized by means of the volume and the area counting, compared with the standard LQC dynamics (the symmetric dashed curve). For both cases the first minimum approximately coincides with the LQC's one $v\approx 126$ while the first maximum differs by slightly less then an order of magnitude, being $v\approx 3165\,$ and $v\approx 371\,$, respectively.}\label{vplot}
\end{figure}
The numerical study has been done with the software Matlab, using a 4th order Runge-Kutta-Merson method for solving the three sets of eqq. of motion corrisponding to the three aforementioned Hamiltonians, with common initial conditions $v_0=10^4,\,b_0=0.005\,$. According to definitions (\ref{deltaLQG}),(\ref{deltaQRLG}),(\ref{deltaVOL}) and (\ref{deltaAREE}) the following values for the area gaps has been used $\tilde{\Delta}=4.78,\,\Delta'=6.03,\,\Delta_{LQG}=5.22\,$.

In fig.\ref{vplot} we see that going backward in time, the volume of the universe follows the LQC standard dynamics until it reaches the first local minimum, i.e. the LQC bounce, and later a departure from the LQC symmetric evolution arises in both cases. A period of damped oscillations follows and the volume settles down to a minimum whose value is close to the value it had at the LQC bounce.

\begin{figure}[H]
\centering
\includegraphics[width=18cm]{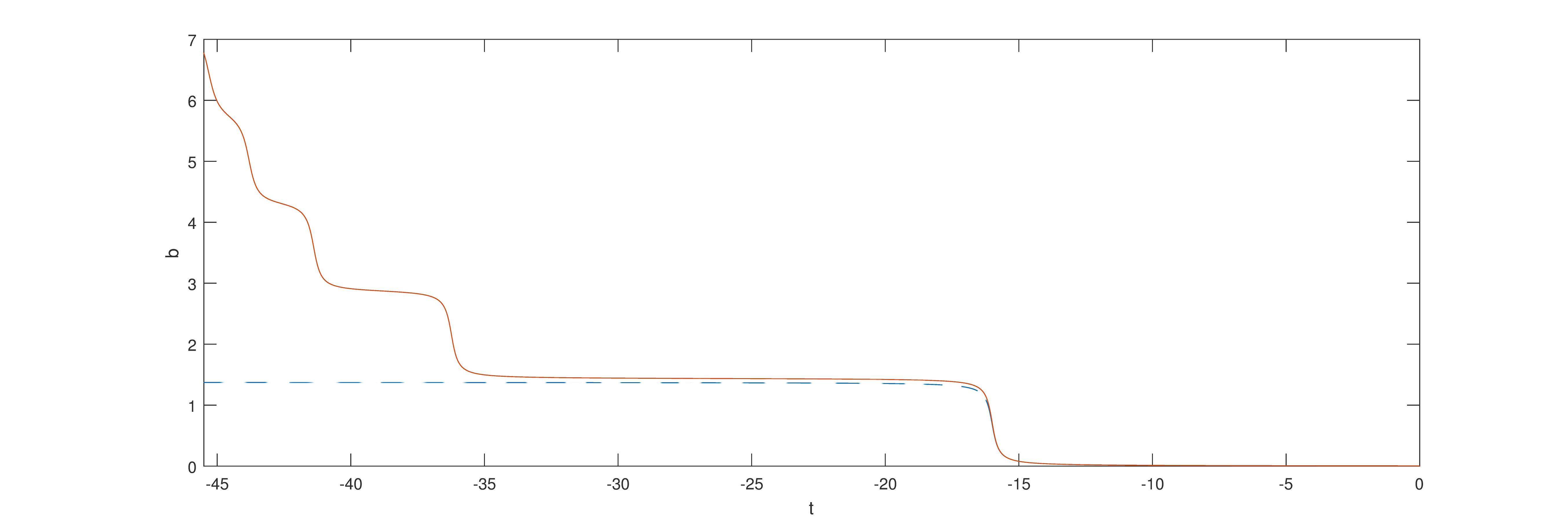}
\includegraphics[width=18cm]{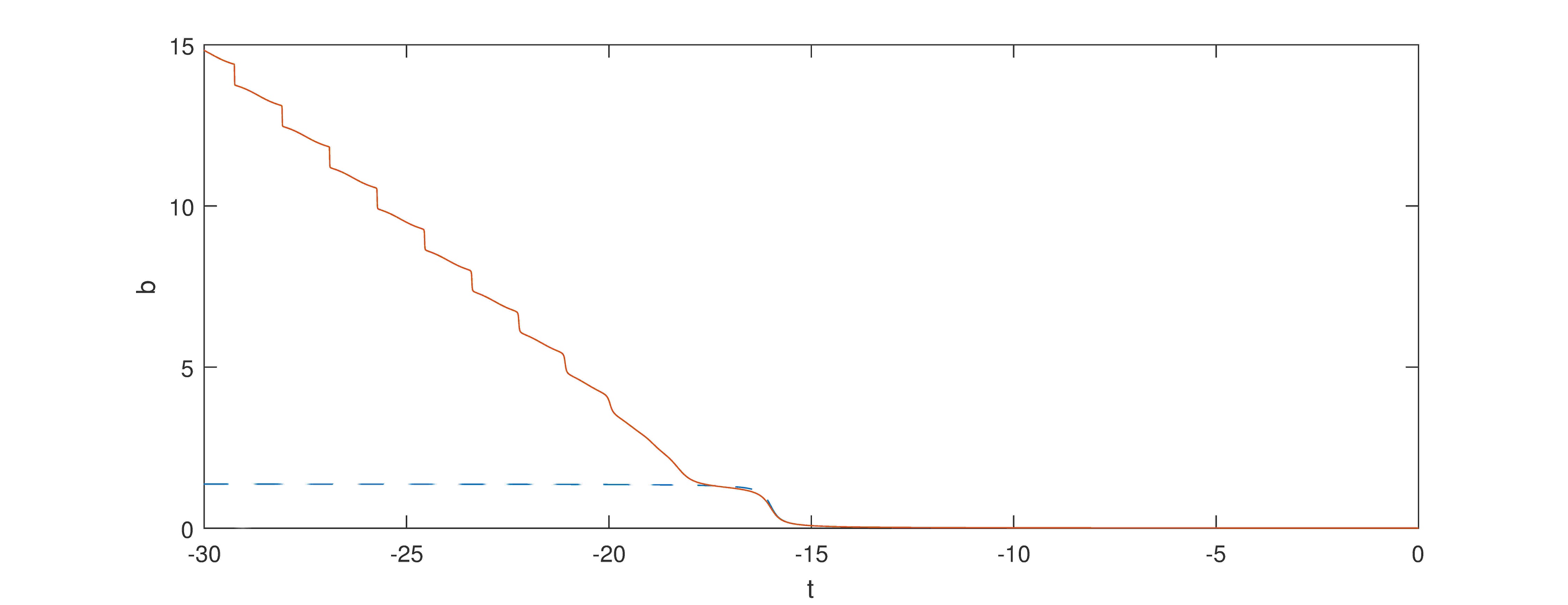}
\caption{From top to bottom the evolution of b vs t according to the QRLG effective Hamiltonians regularized by means of the volume and the area counting, compared with LQC solution (the dashed curve in both the plots). The value of b at the LQC bounce ($t\approx -16$) is $\approx 0.72$ and $\approx 0.63$ in the two cases respectively. Note that going backward in time, the area counting b grows faster then it does in the volume counting case.}
\label{bplot}
\end{figure}

The second plot of fig.\ref{vplot} has a shorter range of times, this is because we have plotted the results of the numerical evolution only for times such that the contribution coming from the terms that are next to the next to the leading order are negligible. Furthermore, we note the dynamics associated to the area counting regularization is more damped, reaching faster its minimum. This can be easily understood realizing that the saddle point approximation is nothing but a power series in which the ratio of each contribution over the next to the leading order go like $v^{-1/3}b$ and $v^{-1/2}b$, respectively for the area counting case and the volume one. Having the former a greater power of $v$ (and also a b growing faster (backward) in time, see fig.\ref{bplot}) the corresponding ratio reaches the order of unity (where we have cut the plot) faster then in the latter case. For the same reason we understand why for early times both the QRLG solutions are indistinguishable from the one provided by LQC, indeed the ratios of the next to the leading orders over the leading (i.e over the first term in the rhs of both (\ref{ciao2}) and (\ref{HFriedareebev})) go like $v^{-2/3}b$ and $v^{-1}$ respectively, thus they vanish when b approaches zero and $v$ bigger values.

\section{Conclusions}

LQG stands as a promising thory for the description of the quantum gravity regime of nature. However, in order to promote its status to the one of \textit{physical} theory, making predictions in agreement with observations is needed. As well know, for quantum gravity this constitutes a tremendous challenge, not yet achieved by any of the tentative theories today at disposal. Anyway, under certain simpler situations, as the one offered by symmetry reduced models, computations leading to predictions can be made and in this perspective LQC deserves here a special mention. But actual computations always require a regularization of the main objects involved in the machinery behind any model.

In this paper we have shown how the $\mu_0$ and $\bar{\mu}$ regularization schemes employed in LQC both comes as special cases of a more natural regularization in QRLG. We have called this regularization {\itshape statistical} because it is based on states prepared in a statistical superposition of graphs: given a probability density distribution for counting the occurence of microstates associated to a fixed macrostate, an effective Hamiltonian can be computed. This procedure also allows to relax the quantization ambiguity associated to the choice of the area gap that is now naturally encoded in the expectation value of the area operator over the chosen density matrix. This point of view leaves room for the proposal of changing the definition of the area gap needed to realize a symmetric bounce for the interior of black holes \cite{Olmedo:2017lvt}.

The first QRLG effective Hamiltonian for the FRLW universe was computed in \cite{Alesci:2016rmn} and studied in \cite{Alesci:2016xqa} where it has been shown it leads to an emergent-bouncing universe, here we have introduced a different counting - an area counting - that allowed us to compute the effective Hamiltonian also for the Bianchi $I$ case. We have applied this new counting also to the Friedmann universe, obtaining an hamiltonian that is different from the one in \cite{Alesci:2016rmn}. As a first result, at the leading order, both the effective hamiltonians computed within the statistical regularization scheme coincide (up to an irrelevant redefinition of the area gap) with the corresponding $\bar{\mu}$ LQC effective ones. Next to the leading order corrections lead to a dynamics that is different from the one provided by LQC.

In the last section we have focussed on the Friedmann case, studying the dynamics generated by the new counting and comparing it both to the one provided by the old counting and LQC. As a main result, numerical investigation has shown that the dynamics associated to the area counting replaces the symmetrical bounce scenario with an evolution that is quasi stationary in the pre-bounce phase and then agrees with the LQC one, supporting the conjecture for the emergent bouncing universe to be a general feature of the isotropic sector of QRLG.  

Finally, interesting questions still deserve to be addressed, like the study of the dynamics generated by the effective QRLG Bianchi $I$ Hamiltonian (does the standard quantum mixmaster scenario radically change?) and the computation of realistic observables, like the power spectrum of perturbations propagating on the effective emergent-bouncing background. In either cases, a relaxation of the saddle point approximation is needed for trusting the evolution also far from the (last) bounce. 

\section*{Acknowledgements}

We wish to thank dr. Giovanni Luzi for tutoring us in the developement of the numerical simulations using the software Matlab and prof. Stefano Liberati for useful discussions during the preparation of the manuscript. EA wish to acknowledge the John Templeton Foundation for the supporting grant \#51876. 


\end{document}